Epitaxial Ba$_2$IrO$_4$ thin-films grown on SrTiO$_3$ substrates by pulsed laser deposition

J. Nichols,[a)] O. B. Korneta, J. Terzic, G. Cao, J. W. Brill, and S. S. A. Seo

*Department of Physics and Astronomy, University of Kentucky, Lexington, KY 40506, USA*
*Center for Advanced Materials, University of Kentucky, Lexington, KY 40506, USA*

**Abstract**

We have synthesized epitaxial Ba$_2$IrO$_4$ (BIO) thin-films on SrTiO$_3$ (001) substrates by pulsed laser deposition and studied their electronic structure by *dc*-transport and optical spectroscopic experiments. We have observed that BIO thin-films are insulating but close to the metal-insulator transition boundary with significantly smaller transport and optical gap energies than its sister compound, Sr$_2$IrO$_4$. Moreover, BIO thin-films have both an enhanced electronic bandwidth and electronic-correlation energy. Our results suggest that BIO thin-films have great potential for realizing the interesting physical properties predicted in layered iridates.

[a)] E-mail: john.nichols@uky.edu

The coexistence of strong spin-orbit coupling and electron-correlation in 5$d$ transition metal oxides has recently attracted lots of attention due to their potential for unprecedented electronic states. For example, a layered iridate compound, Sr$_2$IrO$_4$ (SIO), which is an antiferromagnetic (T$_N$ ~ 240 K) insulator,[1,2] has been proposed as a $J_{eff}$ = 1/2 Mott insulator.[2,3] Its electronic as well as structural similarities to La$_2$CuO$_4$, a parent compound to high-$T_c$ superconductors, have led to a theoretical prediction of unconventional high-$T_c$ superconductivity[4,5] in this layered iridate system. Moreover, due to strong spin-orbit coupling, it is expected to exhibit physical properties that are governed by their topological nature (e.g. Weyl semimetals).[6-8] However, the $J_{eff}$ = 1/2 Mott insulator picture has recently been challenged by SIO being proposed to be a Slater insulator.[9-11] Although there have been some experimental and theoretical efforts on SIO, this remains an open issue. Hence, understanding the true ground state of layered iridate compounds is a very important task since it will direct us to what physical properties should be pursued in future studies.

Ba$_2$IrO$_4$ (BIO) is another layered iridate compound available for us to unveil the physics of coexisting strong spin-orbit coupling and electron-correlation. BIO is also an antiferromagnetic (T$_N$ ~ 240 K) insulator,[12] and angle resolved photoemission spectroscopy[13] as well as X-ray resonant magnetic scattering[12] on BIO show that its electronic and magnetic structure is quite isomorphic to SIO. However, there is a noticeable structural difference between SIO (space group I4$_1$/*acd*) and BIO (space group I4/*mmm*) in their Ir-O-Ir bond angles of 157° and 180°, respectively, which means the IrO$_6$ octahedra are not rotated in BIO.[14] Recently, pressure-dependent experiments on undoped[15] and doped[16] BIO samples have revealed a metallic state. Note that the metallic state is not observed in SIO since the insulating nature of SIO is quite robust.[17-20] Hence, investigations of BIO have been thought to be a promising way of tuning the physical properties of layered iridate compounds in order to ultimately reveal the theoretically predicted properties in the iridates. However, synthesizing BIO crystals is a formidable task due to the fact that it requires a sophisticated high-pressure (P > 6 GPa) synthesis technique.[14] Experimental studies on BIO have been limited to polycrystals[14] or tiny single crystal[13] samples so far. Hence, there is a demand for large area single-crystal and thin-film samples for the wide range of experimental studies and the potential device applications anticipated from this material.

In this letter, we report that epitaxial BIO thin-films can be grown on $SrTiO_3$ (STO) substrates by pulsed laser deposition. The high-pressure conditions required for the synthesis of BIO has been overcome via compressive strain from the substrates. The *dc*-transport and optical spectroscopic data show that the epitaxial BIO thin-films have smaller bandgap energy, wider electronic bandwidths, and more enhanced effective electron-correlation energy than SIO thin-films. Our results provide another technique to fabricate epitaxial thin-films of compounds that require high-pressure-synthesis and to investigate their physical properties.

The epitaxial BIO thin films are deposited on STO (001) single crystalline substrates with a custom built pulsed laser deposition system.[21] The films are approximately 10 nm thick and the preparation of the atomically-flat STO surfaces is described in Ref. 22. The growth conditions are the following: an oxygen partial pressure of 10 mTorr, substrate temperature of 700 °C, and laser (KrF excimer, $\lambda = 248$ nm) fluence of 1.2 J/cm$^2$. Bulk BIO has lattice parameters of $a = 4.030$ Å and $c = 13.333$ Å.[14] The lattice mismatch between bulk BIO and STO is −3.2%, resulting in in-plane compressive strain on the BIO thin-film. Assuming the Young's modulus (Y) of BIO to be about 300 GPa, the in-plane pressure (P) of approximately 9 GPa (P = Y·$\varepsilon_{xx}$) would be exerted on the BIO films by the compressive strain ($\varepsilon_{xx}$), which satisfies the high pressure conditions required for the synthesis of bulk BIO. Polycrystal targets have been synthesized by conventional solid-state sintering and annealing processes using $BaCO_3$ and $IrO_2$ powders. Since high-pressure synthesis techniques are not used, the poly-crystal target consists of various phases of barium-iridium oxide composites with the appropriate Ba:Ir ratio of 2:1, which is confirmed through energy-dispersive X-ray spectroscopy.

The structural properties of the epitaxial BIO thin-films are measured with X-ray diffraction (XRD) using a Bruker D8 Advance system with Cu-Kα radiation. Figure 1 (a) shows an XRD $\theta$-$2\theta$ scan with the (00$l$) peaks of the BIO thin-film. X-ray reciprocal space mapping (RSM) near the STO (103)-plane (Fig. 1 (b)) shows that the (109)-plane of BIO thin-films with a slight strain relaxation (dashed line), which is common for thin films with such a large lattice mismatch. The green asterisk (∗) represents the (109)-plane of bulk BIO. The average lattice parameters of the BIO thin-film estimated from the RSM are $a = 3.91$ Å and $c = 13.45$ Å, which correspond to crystal strains of $\varepsilon_{xx} = -3.0$ % and $\varepsilon_{zz} = +0.8$ % and a Poisson ration of $\nu = 0.12$. The small Poisson ration ($\nu < 0.33$) implies that the BIO thin-film does not sufficiently elongate along the $c$-axis under the in-plane compressive strain. This behavior has also been observed in compressively strained SIO thin-films.[19] The rocking curve scan of the BIO (006) peak (Fig. 1 (c)), whose full-width half-maximum (FWHM) is 0.07°, confirms the good crystallinity of the BIO thin-films. For comparison, the FWHM of the STO (002) rocking curve peak is 0.06° (data not shown). The four-fold symmetry of the BIO thin-film is also confirmed by pole-figure scans of the BIO (103) reflection (Fig. 1 (d)).

Transport measurements (Fig. 2) show that the samples are insulating with a temperature-dependent energy gap estimated from the activation energy ($\Delta_{res} = 2E_a$). The temperature dependence of the resistivity is shown in Fig. 2 (a) for both BIO (red) and SIO (blue) thin-films grown on STO substrates. An Arrhenius plot $\left(\rho = \rho_0 e^{\Delta_{res}/2k_BT}\right)$ is presented in Fig. 2 (b), where $k_B$ is the Boltzmann constant. The magnitude of $\Delta_{res}$ is estimated at two temperature regions for both samples and is smaller for BIO thin-films than for SIO thin-films at all temperatures. It is also noteworthy that $\Delta_{res}$ for both BIO and SIO thin-films decrease as temperature decreases. This abnormal temperature-dependence of gap energy suggests that they become less insulating at low temperature and has also been observed in iridate bulk-crystals[23] and thin-films.[18] This indicates that the insulating nature of this system is quite distinct from simple band insulators.

Optical spectroscopic measurements show that BIO thin-films have a similar electronic structure to SIO thin-films with a few different features. The optical absorption coefficient spectra are presented in Fig. 3 for BIO (a) and SIO (b) thin-films. While both spectra have a qualitatively similar shape, the optical transition peaks (α and β) of BIO thin-films are broader and at higher energies than those of SIO. This means the electronic bandwidth (*W*) is larger for BIO thin-films than for SIO thin-films. Although α and β occur at higher energies, the significantly larger *W* results in the optical gap energy, as estimated from the onset of the optical spectra, of BIO thin-films being smaller than not only SIO thin-films deposited on STO, but for bulk[24] and other thin-film[18,19] samples of SIO as well.

The gap energies of BIO thin-films estimated through *dc*-transport and optical spectroscopic experiments are consistently smaller than SIO thin-films. The increased *W*, which implies the hopping integral (*t*) is larger, reduces the gap energies presumably due to the increased Ir-O-Ir bond angle in the BIO thin-films. However, complete gap closing (i.e. a metallic state) does not occur in this system which remains insulating. We can find an important clue about the insulating nature of BIO thin-films from their optical spectra. As mentioned above, α and β of BIO thin-films occur at higher photon energies than for SIO thin-films. Since the position of these optical transitions is related to $U_{eff}$,[24,25] the increased optical-transition peak positions imply that $U_{eff}$ is also larger in the BIO thin-films. An enhanced $W$ and $U_{eff}$ has been observed in tensile-strained[19] and *a*-axis oriented[18] SIO thin-films as well. At this moment, it is difficult to understand how $U_{eff}$ can be increased by changes in the lattice of layered iridates, since it requires microscopic studies of local ionic and electronic structure.

It is remarkable that even though $U_{eff}$ is larger, the electronic structure is dominated by $W$ broadening, which results in BIO thin-films having a significantly reduced optical gap energy. Note that the hydrostatic pressure-induced metal-insulator transition has been observed in bulk BIO,[15] but not in SIO,[17] which suggests that the electronic structure of BIO is close to the edge of the metal-insulator phase transition. One of the remarkable theoretical predictions for layered iridate compounds is for superconductivity to be realized with carrier doping.[4,5] However, superconductivity has not been observed in SIO samples even under various physical tuning parameters such as electrochemical doping,[23,26,27] hydrostatic pressure,[17] and lattice strain.[19] Our experimental observations on BIO thin-films suggest that BIO is a better candidate for intriguing transport properties such as unconventional superconductivity since its electronic structure is closer to the metal-insulator transition due to significantly reduced transport and optical gap energies.

In summary, we have successfully grown epitaxial BIO thin-films on STO substrates by pulsed laser deposition. By transport and optical spectroscopic measurements, we have observed that the BIO thin-films are still insulating but with an appreciably smaller energy gap than SIO. While they have similar electronic structure in character, the BIO thin-films show a larger bandwidth and effective electronic-correlation energy than SIO. We suggest that the BIO thin-films have great potential for unveiling the intriguing physical properties predicted in the layered iridate compounds due to their electronic structure being close to the metal-insulator transition.

**Acknowledgements**

This research was supported by the NSF through Grant Nos. EPS-0814194 (the Center for Advanced Materials), DMR-1262261 (JWB), DMR-0856234 (GC), DMR-1265162 (GC), and by the Kentucky Science and Engineering Foundation with the Kentucky Science and Technology Corporation through Grant Agreement No. KSEF-148-502-12-303 (SSAS).

**Figure Captions**

**Figure 1** X-ray diffraction of the BIO films on STO. a) A $\theta$-$2\theta$ scan where the BIO peaks are clearly labeled and the STO peaks are identified with a ▼ symbol. b) A reciprocal space map near the STO (103) peak, where the dashed line and the green asterisk (∗) represent the position of the substrate peak and bulk lattice parameters, respectively. c) A rocking curve about the BIO (006) peak ($\omega = 20.33°$). d) A pole figure about the BIO (103) peak in the range of psi from 0° to 60°.

**Figure 2** a) Temperature dependence of the normalized resistivity for BIO (red) and SIO (blue) thin-films on STO substrates. b) Arrhenius plot with gap energy ($\Delta_{res} = 2E_a$) estimated for two temperature regions for BIO (red) and SIO (blue). Note that SIO curve has a vertical offset for clarity.

**Figure 3** Optical absorption coefficient spectra for a) BIO and b) SIO. Inset: Schematic drawing of the electronic structure of the Ir $5d$ band for a) BIO and b) SIO.

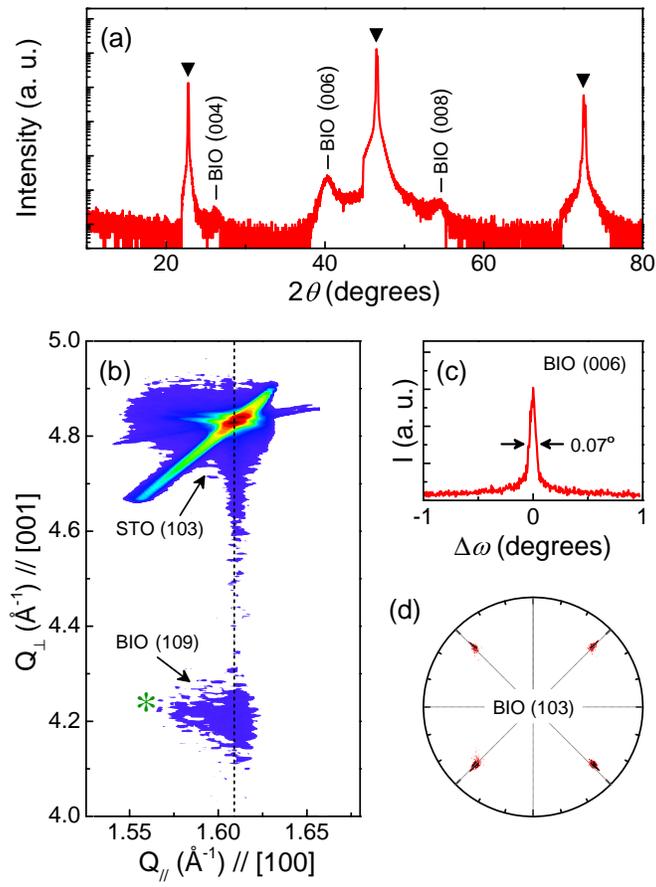

Nichols *et al.*

Figure 1

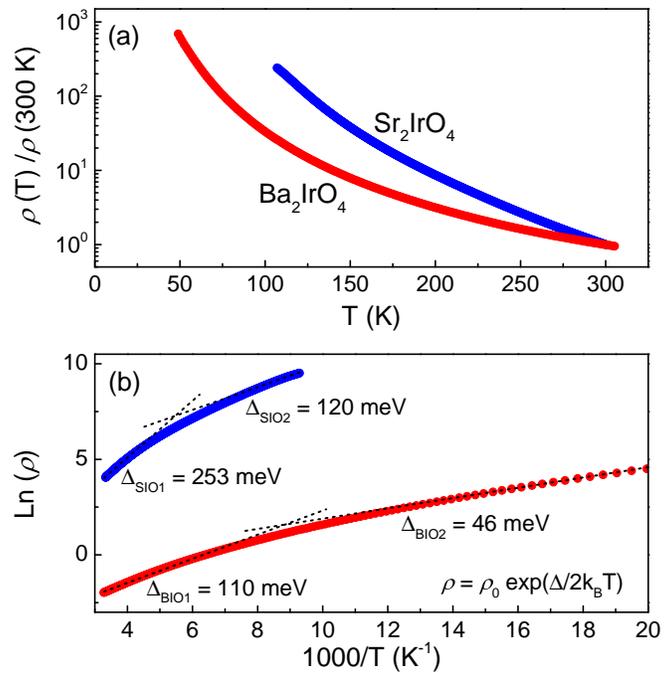

Nichols *et al.*
Figure 2

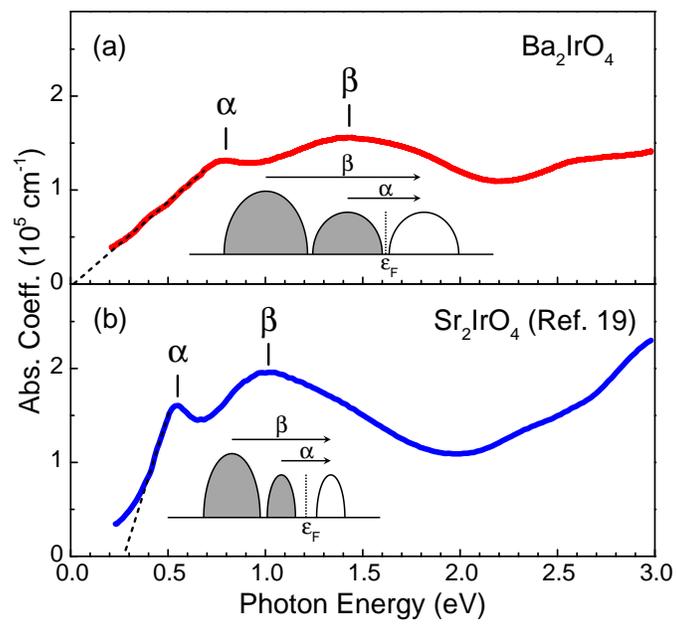

Nichols *et al.*
Figure 3